\documentclass{article}

\usepackage[nonatbib,final]{neurips_2021}
\usepackage[square,numbers]{natbib}

\usepackage[utf8]{inputenc} 
\usepackage[T1]{fontenc}    
\usepackage{hyperref}       
\usepackage{url}            
\usepackage{booktabs}       
\usepackage{amsfonts}       
\usepackage{nicefrac}       
\usepackage{microtype}      
\usepackage{xcolor}         
\usepackage{amssymb}
\usepackage{amsmath}
\usepackage{physics}
\usepackage{graphicx}
\usepackage{color}
\usepackage{bbm}
\usepackage{rotating}
\usepackage{ulem}

\DeclareMathOperator{\id}{id}

\author{
Di Luo \thanks{Co-first authors} \quad  \thanks{Corresponding author: diluo@mit.edu} \\
Center for Theoretical Physics, \\
Massachusetts Institute of Technology,\\ 
Cambridge, MA 02139, USA \\
Department of Physics, Harvard University, \\Cambridge, MA 02138, USA \\
The NSF AI Institute for Artificial \\ Intelligence and Fundamental Interactions \\
\And
Shunyue Yuan $^{*}$ \\
Department of Physics, \\
University of the Illinois, Urbana-Champaign \\
Urbana, IL 61801, USA \\
Department of Applied Physics \\ and Materials Science, \\
California Institute of Technology \\
Pasadena, CA 91125, USA \\
\And
James Stokes \\
Department of Mathematics, \\
University of Michigan, Ann Arbor \\
MI 48109, USA \\
\And
Bryan K. Clark \\
Institute for Condensed Matter Theory and IQUIST and \\
NCSA Center for Artificial Intelligence Innovation \\ and 
Department of Physics, \\
University of the Illinois, Urbana-Champaign \\
Urbana, IL 61801, USA \\
}

\begin{document}
\hspace*{\fill}MIT-CTP/5489
\title{Gauge Equivariant Neural Networks for 2+1D U(1) Gauge Theory Simulations in Hamiltonian Formulation}

\date{\today}

\maketitle
\begin{abstract}

Gauge Theory plays a crucial role in many areas in science, including high energy physics, condensed matter physics and quantum information science. In quantum simulations of lattice gauge theory, an important step is to construct a wave function that obeys gauge symmetry. In this paper, we have developed gauge equivariant neural network wave function techniques for simulating continuous-variable quantum lattice gauge theories in the Hamiltonian formulation. We have applied the gauge equivariant neural network approach to find the ground state of $2+1$-dimensional lattice gauge theory with U(1) gauge group using variational Monte Carlo. We have benchmarked our approach against the state-of-the-art complex Gaussian wave functions, demonstrating improved performance in the strong coupling regime and comparable results in the weak coupling regime.
\end{abstract}

\section{Introduction}

Parametrized gauge equivariant functions play an increasingly important role in deep learning and the quantum simulation literature.
The historical importance of gauge equivariance in physics literature originates from the gauge principle, according to which the action functional dictating the interactions of fundamental fields must be invariant under symmetry transformations that depend upon $d+1$-dimensional space-time. It follows from the gauge principle that the form of the action functional is fixed in terms of equivariant features of the field arguments, up to a low-dimensional undetermined parametrization called coupling parameters. Remarkably, the gauge principle is sufficient to essentially completely determine the form of the quantum Hamiltonian $H$, leaving the determination of its eigenmodes as a problem of gauge invariant function approximation. 
This paper offers a new approach to the ground eigenvalue problem for a prototypical gauge theory in $2+1$ dimensions known as the Kogut-Susskind model, by drawing on techniques from gauge equivariant machine learning. In contrast to an existing literature which pursues Lagrangian formulation~\cite{kanwar2020equivariant}, this paper works in the so-called Hamiltoinan lattice gauge theory, in which the spatial manifold is described by a $d$-dimensional regular periodic grid graph $(V,E)$ with vertices $V$ and edges $E$, and the space of fields is described by a high-dimensional manifold $\Omega = \prod_{e \in E} G$ given by the product of some compact Lie group $G$ over the edges of the graph. The quantum Hamiltonian $H$ is defined by a Schr\"{o}dinger operator acting on the space of wave functions 
$L^2(\Omega) = \bigotimes_{e \in E} L^2(G)$.
A  gauge transformation is represented by an element $g\in \prod_{v\in V}G$, which acts locally on a given field configuration
$x \in \Omega$ by conjugation, producing a transformed field configuration, which we denote by $g\cdot x \in \Omega$.
The gauge group correspondingly acts wavefunctions $\psi \in L^2(\Omega)$ as $g\cdot\psi(x) = \psi(g^{-1}\cdot x)$ and the quantum Hamiltonian is equivariant with respect to the gauge group in the sense that $g\cdot(H\psi) = H(g\cdot\psi)$ for all $g\in \prod_{v\in V}G, \psi \in L^2(\Omega)$. For the ground states of pure gauge theories with no matter, the wave functions satisfy the strict invariance property $g\cdot\psi = \psi$.

\begin{figure}[t]
    \centering
    \includegraphics[scale=0.3]{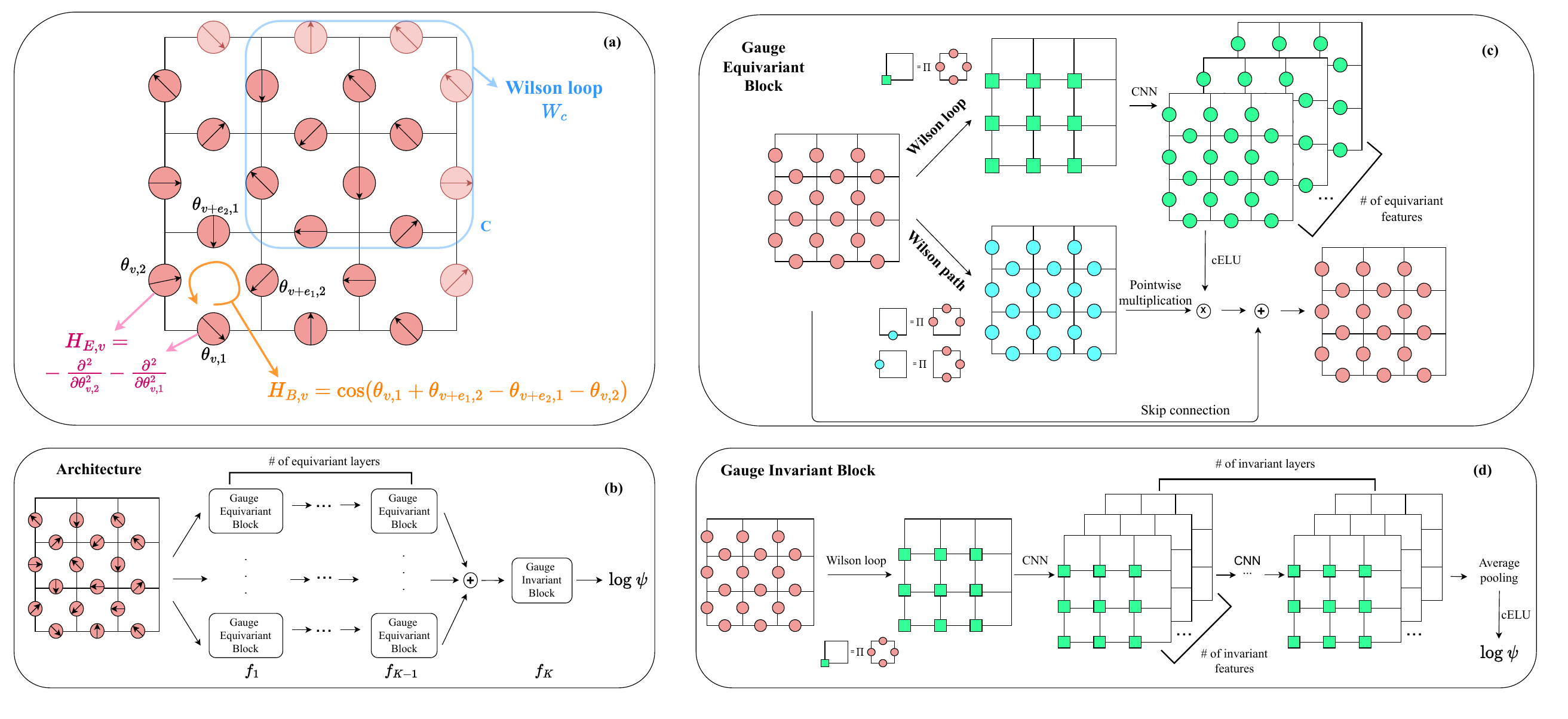}
    \caption{(a) Hamiltonian of U(1) lattice gauge theory. 
    The gauge field is defined on each edge by a set of angular variables $\theta$ between $[0,2\pi)$ and represented as $e^{i\theta}$.
    The Hamiltonian is defined in Eq.~\ref{eq:ham}, where $H_{E,v}$ describes the kinetic energy of the gauge field for the vertex $v$ and $H_{B,v}$ describes the magnetic energy. The faded arrows on edges are used to represent periodic boundary condition. 
    (b) Gauge equivariant neural network architecture.
    (c) Gauge equivariant block. 
    (d) Gauge invariant block.
    }
    \label{fig: architecture}
\end{figure}

In general, a wave function lives in a high dimensional Hilbert space which dimensionality grows exponentially with the number of particles, which is one of the main challenges for solving quantum many-body physics. With the advancement of machine learning, neural networks have been proposed ~\cite{Carleo602} to efficiently represent quantum wave functions, which is known as neural-network quantum states (NNQS). The key idea is to find a compact neural network representation of the high dimensional quantum state. It has been shown recently NNQS is able to approximate a large family of quantum state efficiently~\cite{sharir2021neural} and any quantum state supervised learning is also guaranteed to converge in the infinite width limit~\cite{luo_inf_nnqs}. In addition, NNQS also provides exact representations for many quantum states ~\cite{Gao2017,topo_wf,Levine_2019,luo_gauge_inv,luo2021gauge,dongling, huang_chiral,Vieijra_2020}. Recent research has applied NNQS to a variety of fields including condensed matter physics~\cite{Han_2020,ferminet,Choo_2019,rnn_wavefunction,Luo_2019,paulinet,wang2021spacetime,Glasser_2018,Stokes_2020,Nomura_2017,py2021}, high energy physics~\cite{luo2021gauge,luo_gauge_inv,nuclear2021}, and quantum information science~\cite{qaoa2021,quantum_circuit,Torlai2018_tomography}. It has been shown that NNQS can achieve state-of-the-art results for computing ground states and the real time dynamics properties of closed and open quantum systems~\cite{gutierrez2020real, Schmitt_2020,Vicentini_2019,PhysRevB.99.214306,PhysRevLett.122.250502,PhysRevLett.122.250501,luo_gauge_inv,luo_povm}.

In this work, we develop the gauge equivariant neural network for U(1) lattice gauge theory. We focus on the U(1) Kogut-Susskind model, which is characterized by the simplest abelian Lie group $G=S^1$ and a $d=2$ dimensional grid graph. We start by embedding $\Omega = \prod_{e\in E}S^1$ in a linear space $\mathbb{C}^{N_1}$ on which the group action is defined, and introduce a sequence of hidden vector spaces $\mathbb{C}^{
N_2},\ldots,\mathbb{C}^{N_K},\mathbb{C}^{N_{K+1}}$ (with $N_{K+1}=1$) and associated nonlinear mappings $f_k : \mathbb{C}^{N_k} \longrightarrow\mathbb{C}^{N_{k+1}}$ such that for all $g \in \Omega$ we have $f_k(g\cdot x) = g \cdot f_k(x)$ for $k \in \{1,\ldots,K-1\}$ and $f_K(g \cdot x) = f_K(x)$ for all relevant configurations $x$. The proposed trial wave function is given by
\begin{equation}\label{e:comp}
    \psi = f_K \circ \cdots \circ f_1
\end{equation}\label{eq:wf_equ}

In early work on Hamiltonian lattice gauge theories, non-compositional trial wave functions have been proposed in which $f_1 = \cdots = f_{K-1} = \id$ as the identity function. Here we have generalized the theory in ~\cite{luo2021gauge} to the U(1) group and constructed continuous-variable gauge equivariant neural network for each $f_k$ with much more powerful representation.

\subsection{Related Work}

In the machine learning community, group equivariant neural network for global symmetries have been considered in ~\cite{max_group,risi_covariant,risi_fourier,steerable_cnn,tess_e3} and gauge equivariant neural networks have been recently introduced in \cite{cohen2019gauge,max_gauge}. The authors in ~\cite{cohen2019gauge} used parallel transport techniques to construct convolutional neural networks defined on manifolds. In the physics community, gauge equivariant neural networks have been utilized to accelerate sampling and observables fitting in Lagrangian formulation of lattice field theories \cite{kanwar2020equivariant,phiala_fermion_flow,phiala_qcd_flow,phiala_sun_flow,Favoni_2022}. Gauge equivariant and gauge invariant neural networks~\cite{luo_gauge_inv,luo2021gauge} have subsequently been used as trial wave functions for Hamiltonian formulation of lattice gauge theories with discrete gauge group including $\mathbb{Z}_N$ theory and non-abelian Kitaev $D(G)$ model. In this work, we advance the state-of-art in \cite{luo2021gauge} by constructing gauge equivariant neural network wave functions for continuous-variable lattice gauge theory and focus on the simplest abelian Lie group U(1).

\section{Theoretical Background}
 
We consider the 2+1D compact U(1) pure gauge theory which is defined on a $L \times L$ square lattice with periodic boundary conditions. There are $V=L^2$ vertices and $2L^2$ edges, where the gauge field is a set of angular variables between $[0,2\pi)$ on each edge. The space of field configurations is then given by $\Omega = [0,2\pi)^{2L^2}$ and we represent a field configuration $x$ by a 2-channel image $x_\delta$ with $\delta=1,2$, which represent the x-axis and the y-axis directions for the gauge field respectively.
\begin{equation}
x_{\delta} \equiv 
\begin{bmatrix}
e^{i\theta_{(1,1),\delta}} & \cdots & e^{i\theta_{(L,1),\delta}} \\
\vdots & & \vdots \\
e^{i\theta_{(1,L), \delta}} & \cdots & e^{i\theta_{(L,L), \delta}}
\end{bmatrix}
\enspace , \quad\quad\quad \delta=1,2
\end{equation}

The Kogut Susskind Hamiltonian is defined as 

\begin{equation}
    H = \sum_{v \in V} \left[-\frac{g^2}{2}  \left(\frac{\partial^2}{\partial \theta_{v, 1}^2}
    +\frac{\partial^2}{\partial \theta_{v, 2}^2}\right)
    - \frac{2}{g^2} \cos (\theta_{v,1} + \theta_{v+e_1,2} - \theta_{v+e_2,1} - \theta_{v,2})\right]\label{eq:ham}
\end{equation}
where the first term describes the kinetic energy of the gauge field, the second term describes the magnetic energy (Fig.~\ref{fig: architecture}(a)) and $g$ is the coupling constant. 

The theory requires an additional satisfaction of gauge symmetry, such that the wave function is invariant under the following transformation on each vertex $v\in V$,
\begin{equation}
    \theta_{v,\delta} \longrightarrow \theta_{v,\delta} + \alpha_{v+e_\delta} - \alpha_v \enspace , \quad\quad\quad \delta=1,2, \quad\quad\quad \alpha_v \in \mathbb{R} \label{eq:trans}
\end{equation}
An important task is to find the ground wave function satisfied the above symmetry for the Hamiltonian in Eq.~\ref{eq:ham}. 

\section{Methods}

\subsection{Model Architectures}
We start our construction of wave functions that satisfies the gauge symmetry in Eq.~\ref{eq:trans} following the spirit of Eq.~\ref{eq:wf_equ}. For each gauge equivariant layer $f_k$ where $l \in \{1,...,K-1\}$ , it has been shown in ~\cite{luo2021gauge} that the construction in Fig.~\ref{fig: architecture}(c) satisfies $f_k(g\cdot x) = g \cdot f_k(x)$ for any $g \in \prod_{v \in V} \mathbb{Z}^N$. The property generalizes to the U(1) group by taking $N \rightarrow \infty$. For the gauge invariant layer $f_k$, since it operates Wilson loop values which are gauge invariant objects,  the construction in Fig.~\ref{fig: architecture}(d) is invariant under the transformation in Eq.~\ref{eq:trans} and satisfies  $f_K(g \cdot x) = f_K(x)$ for $g \in \Omega=\prod_{v \in V} \mathbb{S}^1$. Hence,  $\psi = f_K \circ \cdots \circ f_1$ is invariant with respect to the transformation in Eq.~\ref{eq:trans} and respects the U(1) gauge symmetry.

We have developed a gauge equivariant neural network architecture as Fig.~\ref{fig: architecture} shows. The general architecture of the gauge equivariant neural network is composed of gauge equivariant layers followed by gauge invariant layer which outputs the logarithmic of the wave function amplitude. The activation function in both equivariant and invariant blocks is $\text{cELU}(z) = \text{ELU}(\mathrm{Re}(z)) + i * \text{ELU}(\mathrm{Im}(z))$ where $z$ is a complex variable, and 
\begin{equation}
\text{ELU}(x)=
    \begin{cases}
        x & \text{for $x > 0$} \\
        e^{x}-1 & \text{for $x \leq 0$}. 
    \end{cases}
\end{equation}

The network architecture is specified through the number of equivariant blocks, equivariant layers, equivariant features, invariant layers and invariant features. There are two gauge equivaraint neural networks that we use in this work: Equ-NN and Equ3-NN. Equ-NN has only one gauge equivariant block in each equivariant layer, while Equ3-NN has three equivariant blocks in each layer. For all our simulations in the following sections, we use 2 equivariant layers, 2 invariant layers, and 2 equivariant features for both Equ-NN and Equ3-NN. The number of invariant features depends on $g^2$ with four features being the default except for $g^2=0.5$ and 0.6 on the Equ-NN where we use three features.

\subsection{Optimization Approach}
We search for the ground state through the variational principle
\begin{equation}
 \min_{w}   E = \frac{\braket{\psi_{w}}{H\psi_{w}}}{\braket{\psi_{w}}{\psi_{w}}} = \frac{\int_\Omega \psi_{w}^{*}(x) H\psi_{w}(x) dx}{\int_\Omega \psi_{w}^{*}(x) \psi_{w}(x) dx}
\end{equation}
where $w$ are the parameters of the gauge equivariant neural network. $\langle \cdot| \cdot\rangle$ denotes the $L^2$ inner product for $L^2(\Omega) = L^2\big([0,2\pi)^{2L^2}\big)$. The optimization is performed through variational Monte Carlo, where the gradients can be computed as follows:

\begin{figure}[t]
    \centering
    \includegraphics[scale=0.9]{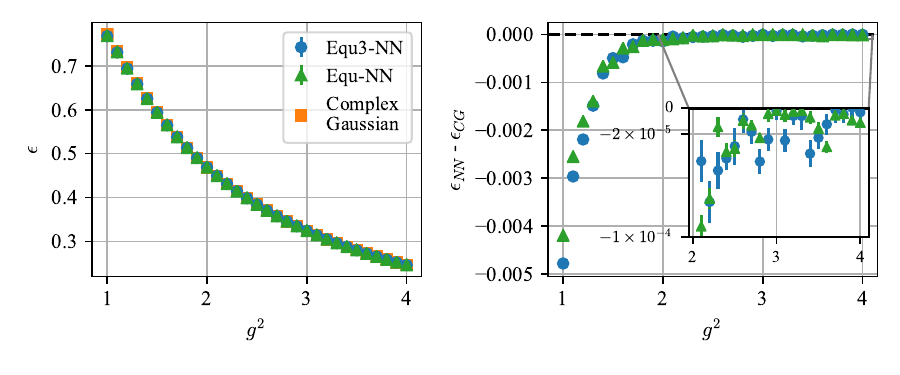}
    \caption{Left: Ground state energies of our variational ansatz compared with the data from \cite{julian2020realtime} for $1 \leq g^2 \leq 4$ at  $L=8$.
    Right: Difference in ground state energies from our variational ansatz ($\epsilon_{NN}$) and the complex Gaussian wave functions ($\epsilon_{CG}$) from  Ref.\cite{julian2020realtime}. Inset shows energy differences in log-scale. }
    \label{fig: largeg_comparison}
\end{figure}

\begin{equation}
    \pdv{w}E \approx \frac{2}{N}\sum_{x \sim \abs{\psi_w}^2}^N \real\left\{E_\text{loc}'(x) \pdv{w} \log \psi_w^*(x)\right\}.
    \label{eq:variational_gradient}
\end{equation}
where
\begin{equation}
    E_\text{loc}'(x) = \frac{H\psi_{w}(x)}{\psi_w({x})}-\frac{1}{N}\sum_{x \sim \abs{\psi_w}^2}^N \frac{H\psi_{w}(x)}{\psi_w({x})}.
\end{equation}
The sampling $x \sim \abs{\psi_w}^2$ is performed through Markov Chain Monte Carlo. Optimization is performed using the Stochastic Reconfiguration method\cite{Sorella2017book}.

\begin{figure}[t]
    \centering
    \includegraphics[scale=0.9]{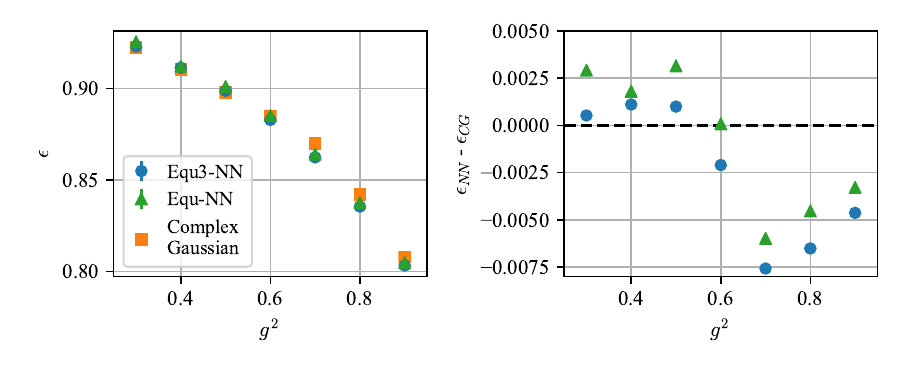}
    \caption{Ground state energy for $0.3 \leq g^2 \leq 0.9$ at $L=14$ measured by gauge equivariant neural networks compared to the complex Gaussian wave functions from  Ref.\cite{julian2020realtime}. Stastical error bars from sampling are plotted within the size of the markers.
    }
    \label{fig: smallg_comparison_lx14}
\end{figure}

\section{Experiments}

We compare our approaches with the state-of-the-art complex Gaussian method for both the strong coupling regime and the weak coupling regime. For the strong coupling regime, $g$ is large and the kinetic energy term is more dominant such that a moderate system size $L$ is sufficient for simulations. We consider $g$ from 1 to 4 with increment 0.1 on $L=8$ systems for experiments. Fig.~\ref{fig: largeg_comparison}(a) shows that both Equ-NN and Equ3-NN produce consistent energy decay as $g$ increases. Fig.~\ref{fig: largeg_comparison}(b) shows the energy difference between the gauge equivariant neural networks and the state-of-the-arts complex Gaussian wave function method~\cite{julian2020realtime}. According to the variational principle, the lower energy indicates better performance. It can be seen that Equ-NN and Equ3-NN are both better than the complex Gaussian wave function and the energy performance gets better as $g$ is closer to 1.

For the weak coupling regime, $g$ is small so that the magnetic energy term in the Hamiltonian of Eq.~\ref{eq:ham} plays a more important role and it requires larger system size to capture the physics. We consider experiments over $g$ from 0.3 to 0.9 with increment 0.1 on $L=14$ systems. Fig.~\ref{fig: smallg_comparison_lx14} shows that the gauge equivariant neural network achieves better energy than the complex Gaussian wave function when $g \geq 0.6$ and comparable results when $g$ is smaller. It is worth notice that in the $g \rightarrow 0$ limit, the ground state wave function becomes the trivial delta function in terms of the plaquette variable basis, which could cause more challenging optimization for the gauge equivariant neural network. It may lead to the difference of the performance compared to the complex Gaussian wave function since a Gaussian is easier to approach a delta function.

To reflect the physics of the theory, one can consider the Wilson loop observable computed by
\begin{equation}
    W_{C} =  \frac{\langle \psi | \cos (\sum_{v \in C}  \theta_{v,1} + \theta_{v+e_1,2} - \theta_{v+e_2,1} - \theta_{v,2}) \psi \rangle}{ \braket{\psi}{\psi}}
\end{equation}
where $C$ is any contour on the lattice as Fig.~\ref{fig: architecture}(a) shows. Since the 2+1D pure U(1) gauge theory is confined for any $g$~\cite{polyakov1975compact}, the Wilson loop should decay exponentially as Eq.~\ref{eq:loop} with respect to the area included by the contour $C$. 
\begin{equation}
    W_C = e^{-\sigma R_1 \times R_2- 2a (R_1+R_2) +c}\label{eq:loop}
\end{equation}
where $C$ is chosen to be a $R_1 \times R_2$ rectangle on the lattice. The exponential decay behaviours of the Wilson loops over $g^2=0.5,0.6,0.7$ of Equ-NN and Equ3-NN are shown in Fig.~\ref{fig: lx14_wilsonloop_Equ-NN} and Fig.~\ref{fig: lx14_wilsonloop_Equ3-NN} in the Supplementary Material.

\section{Conclusion and Discussion}

Gauge theories appear in a variety of contexts in physics, including high energy physics, condensed matter physics and quantum information science, and provides fundamental understandings of our universe. In this work, we have developed gauge equivariant neural networks for continuous-variable quantum lattice gauge theory with U(1) gauge group. We have applied the approach to find the ground state of 2+1D U(1) pure gauge theory and demonstrated better performance in the strong coupling regime and comparable results in the weak coupling regime over the state-of-the-arts complex Gaussian wave function. This is an important step that goes beyond the discrete lattice gauge theory \cite{luo2021gauge} and opens up the possibilities of simulating the ground state and real time dynamics of lattice gauge theories with different symmetry groups. Future research will generalize the approach to non-abelian lattice gauge theories and study important physics such as confinement.  

\section{Acknowledgements}
The authors would like to thank Julian Bender for providing data in Ref.~\cite{julian2020realtime} for comparison, and acknowledge helpful discussion with Giuseppe Carleo, Lena Funcke, and Zhuo Chen. DL acknowledges support from the Co-Design Center for Quantum Advantage (C2QA) and the NSF AI Institute for Artificial Intelligence and Fundamental Interactions (IAIFI).
JS acknowledges support from NSF under grant DMS-2038030. BKC acknowledge support from the Department of Energy grant DOE DESC0020165.

\bibliographystyle{abbrv}
\bibliography{bibo/references}

\clearpage

\renewcommand\thefigure{S\arabic{figure}}  
\renewcommand\thetable{S\arabic{table}}  
\renewcommand{\theequation}{S\arabic{equation}}
\renewcommand{\thepage}{P\arabic{page}} 
\setcounter{page}{1}
\setcounter{figure}{0}  
\setcounter{table}{0}
\setcounter{equation}{0}

\begin{center}
	\noindent\textbf{\large{Supplementary Material}}
\end{center}
\appendix

\begin{figure}[h!]
    \centering
    \includegraphics[scale=0.6]{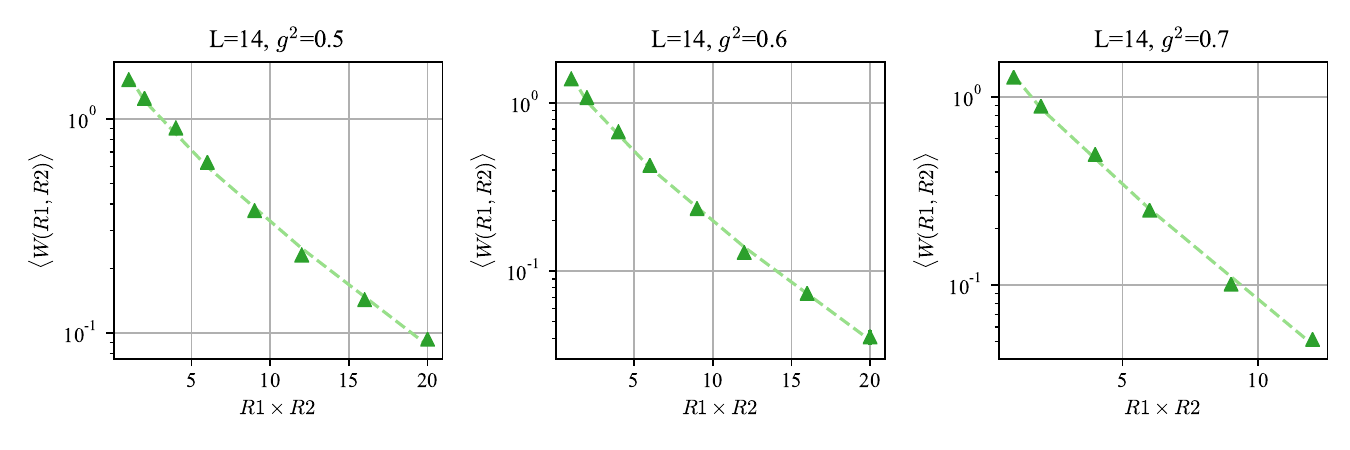}
    \caption{Wilson loops $\langle W(R_1, R_2)\rangle $ in the ground state in the small coupling regime, computed on a $14 \times 14$ lattice by model Equ-NN, as a function of the area $R_1 \times R_2$.  The dash line fittings show the exponential decay of Wilson loops according to Eq.~\ref{eq:loop}. 
    }
    \label{fig: lx14_wilsonloop_Equ-NN}
\end{figure}

\begin{figure}[h!]
    \centering
    \includegraphics[scale=0.6]{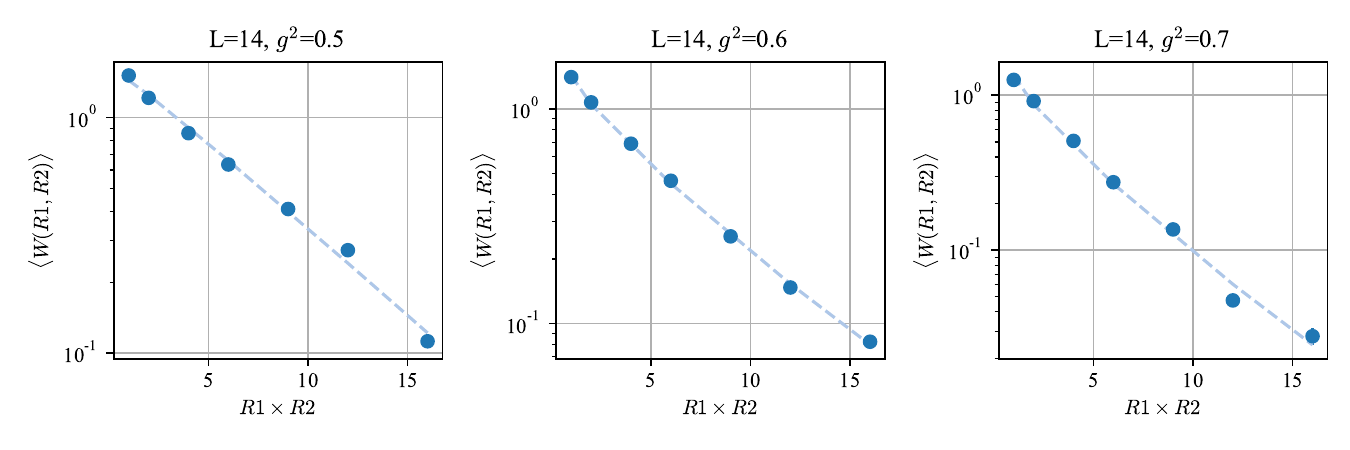}
    \caption{Wilson loops $\langle W(R_1, R_2)\rangle $ in the ground state in the small coupling regime, computed on a $14 \times 14$ lattice by model Equ3-NN, as a function of the area $R_1 \times R_2$.  The dash line fittings show the exponential decay of Wilson loops according to Eq.~\ref{eq:loop}. }
    \label{fig: lx14_wilsonloop_Equ3-NN}
\end{figure}

\end{document}